\documentclass{aa}

\usepackage{graphicx}
\usepackage{txfonts}
\usepackage{natbib}

\bibpunct{(}{)}{;}{a}{}{,} 

\begin{document}

\title{The MeqTrees Software System And Its Use For Third-Generation Calibration Of Radio Interferometers}

\author{J.E.\ Noordam \and O.M.\ Smirnov}

\institute{Netherlands Institute for Radio Astronomy (ASTRON)\\
  P.O. Box 2, 7990AA Dwingeloo, The Netherlands \\
  \email{noordam@astron.nl smirnov@astron.nl}}

\date{Received 18 May 2010 / Accepted 16 August 2010}

\titlerunning{MeqTrees Software System}
\authorrunning{J.E.\ Noordam \& O.M.\ Smirnov}

\abstract%
{The formulation of the radio interferometer measurement equation (RIME) for 
a generic radio telescope by Hamaker et al. has provided us with an elegant
mathematical apparatus for better understanding, simulation and
calibration of existing and future instruments. The calibration of the
new radio telescopes (LOFAR, SKA) would be unthinkable without the RIME
formalism, and new software to exploit it.}%
{The MeqTrees software system is designed to implement numerical
models, and to solve for arbitrary subsets of their parameters. It may
be applied to many problems, but was originally geared towards
implementing Measurement Equations in radio astronomy for the purposes
of simulation and calibration. The technical goal of MeqTrees is to
provide a tool for rapid implementation of such models, while offering
performance comparable to hand-written code. We are also pursuing the
wider goal of increasing the rate of evolution of radio astronomical
software, by offering a tool that facilitates rapid experimentation,
and exchange of ideas (and scripts).}%
{MeqTrees is implemented as a Python-based front-end called the
meqbrowser, and an efficient (C++-based) computational back-end called
the meqserver. Numerical models are defined on the front-end via a
Python-based Tree Definition Language (TDL), then rapidly executed on
the back-end. The use of TDL facilitates an extremely short
turn-around time (hours rather than weeks or months) for
experimentation with new ideas. This is also helped by unprecedented
visualization capabilities for all final and intermediate results. A
flexible data model and a number of important optimizations in the
back-end ensures that the numerical performance is comparable to that
of hand-written code.}%
{MeqTrees is already widely used as the simulation tool for new
instruments (LOFAR, SKA) and technologies (focal plane arrays). It has
demonstrated that it can achieve a noise-limited dynamic range in
excess of a million, on WSRT data. It is the only package that is
specifically designed to handle what we propose to call {{\em third-generation}} 
calibration (3GC), which is needed for the new generation
of giant radio telescopes, but can also improve the calibration of
existing instruments.}%
{}


\maketitle

\section{Introduction\label{sec:Introduction}}

The MeqTrees software system has been designed to implement an
arbitrary Measurement Equation (i.e. a numerical model of an
instrument and/or process), and to solve for arbitrary subsets of its
parameters. In this paper we will concentrate on the simulation
and calibration of data taken with radio telescopes. After all, that
is the subject for which MeqTrees was developed originally, and for
which it is most urgent. It is also an excellent subject for
demonstrating the special capabilities of MeqTrees.

Until recently, radio interferometers like WSRT, VLA, ATCA, GMRT were
designed so that they could be approximated by a relatively simple
instrumental model. Their {\em stations}\footnote{Throughout this
paper, we will use the generic term {\em station} for an element of an
interferometer array. A station can be a parabolic dish or an aperture
array, or something more exotic like a parabolic cylinder. Each
station has two output signals, one for each polarization.} were
carefully designed so that the shape of their spatial response
beams could be assumed to be `identical' at all times. In addition, the
instrumental error associated with a station can be represented in
this model by only two complex gain factors (one for each polarization),
which may vary in time and frequency.  Because their stations are
parabolic dishes that are pointed towards the observed field,
instrumental polarization effects can be treated (or not) as small
`leakage' terms.

Before 1980, first-generation calibration (1GC) was based on
`open-loop' methods, making separate calibrator observations before
and after, and relying on instrumental stability in between. The
result was a dynamic range\footnote{Dynamic range is the ratio between
the flux of the brightest source in the field, and either the thermal
noise or the calibration artifacts, whichever is higher.} of about
100:1.  However modest, this was enough for a plethora of important
discoveries.

Around 1980, the invention of {\em self-calibration} \citep[see also summary by \citealt{Ekers:Selfcal}]{Cornwell:selfcal} ushered in the era of second-generation
calibration (2GC). Selfcal is a ``closed-loop'' method which
continuously estimates the complex station gain factors with the help
of one or more bright sources in the field of view. In this process,
the utilized Sky Model is improved also.  Selfcal has been
spectacularly successful, and has led to a blossoming of techniques,
software packages, and beautiful results. It allows astronomers to
achieve dynamic ranges in excess of $10^4-10^5$ as a matter of
routine, depending on how well the instrument approximates its
simplified model. Record dynamic ranges of well over $10^6$ have been achieved at the WSRT by
\citet{deBruyn:million,deBruyn:3C147} (see also Sect. \ref{sec:Results}).

Radio astronomy is currently going through a remarkable worldwide
burst of building new telescopes and upgrading existing
ones\footnote{An important incentive is the preparation for the
building of the multi-billion Euro Square Kilometer Array (SKA) later in this
decade.}. These instruments present a new, two-pronged calibration
challenge. On the one hand, they are much more sensitive, so more
subtle instrumental effects will have to be taken into account to
reach the thermal noise. On the other hand, the use of new technology
like phased arrays complicates the instrumental model. Therefore, what we propose to call 
{\em third-generation calibration} (3GC) will require a more complex,
able, and a general form of selfcal.

In 1996, \citeauthor{ME1} developed a formulation of an explicit 
radio interferometer measurement equation (RIME) for a generic radio telescope. 
Further work by \citet{ME4} led to a fully $2\times2$
matrix formulation of the RIME, which provides the mathematical
underpinnings for 3GC. Without this full-polarization formalism,
calibration of the new telescopes would be difficult. However,
although the RIME is widely recognized as being correct, complete and
universal, its actual adoption has been slow.  This is caused to a
large extent by the sustained success of the existing 2GC data
reduction packages (AIPS, NEWSTAR, MIRIAD, DIFMAP). The ensuing low
rate of evolution of calibration techniques could be a risk factor for
the new telescopes.

One of the most important challenges of 3GC is dealing with
Direction-Dependent Effects (DDEs), i.e. instrumental effects that can
no longer be assumed to be constant over the field of view. The most
important DDEs are typically caused by the ionosphere (mostly phase
and Farady rotation), and by station beamshapes that differ
substantially from each other, and/or vary individually in frequency
and time. Tackling DDEs implies that one has to solve for a much
larger number of RIME parameters than before.  Besides the practical
problem of extra processing (which may well turn out to be a major
bottleneck, but will not be discussed here), this raises some more
fundamental issues about whether there is enough information available
for 3GC. And, last but not least, methods are needed to correct for
DDEs once they are known, which is non-trivial.

At this moment, it is not yet clear how some of the new generation of
telescopes will be calibrated to the necessary precision. The MeqTrees
software system is a tool that can play a role in building that understanding
on several levels.  First of all, it is firmly based on the explicit RIME. 
Secondly, it has many built-in features for {\em generalized selfcal}, such as
allowing for arbitrary RIMEs with arbitrary parameterizations, and
solving for arbitrary subsets of RIME parameters, including source
parameters. These parameterizations may be experimented with rather rapidly, since
the art of modelling (in Python) is separated from the complex and efficient numerical 
machinery (in C++) ``under the hood''. Rapid progress is also greately helped by
the many possibilities for visualization of
intermediate results, enabling one to easily see what is actually going
on.

The heart of this paper, Sects.~\ref{sec:forest}--\ref{sec:performance}, is a detailed description of how
MeqTrees works. Sect.~\ref{sec:RIME} gives a brief description of the RIME, explains why it is such a powerful 
formalism, and shows how it pertains to MeqTrees. In Sect.~\ref{sec:Results}, we present some recent results that give a taste of what MeqTrees can do.

\section{\label{sec:forest}Nodes, trees, forests}

This section gives a broad overview of MeqTrees design (Sect.~\ref{sec:Design}) and implementation (Sect.~\ref{sec:Implementation}). Subsequent sections will then elaborate on some of the important concepts and features.

\subsection{Design: Expression Trees\label{sec:Design}}

According to Donald Knuth, ``trees have been in existence since the
third day of creation, and perhaps earlier.'' The use of {\em trees}
as information structures is ubiquitous throughout computing science;
\citet{Knuth:Trees} provides a good introduction. MeqTrees uses trees
to represent mathematical expressions. This particular idea dates back
to very early work on compilers \citep{Hopper:Trees}, and in fact was
the first application of trees to computer science.

A tree is a type of graph whose {\em nodes} are connected in a parent-child
hierarchy. A tree node can be parent to zero or more child nodes, and
can itself be a child of zero or more other nodes.\footnote{By the strict conventional definition of
a tree graph, a node may have at most one parent. MeqTrees allows for multiple parents; the proper term for such an
structure is {\em directed acyclic graph}. We use {\em tree} for brevity.} 
Cycles are not allowed. A node having no children is traditionally called a {\em
leaf}, a node with no parents is called a {\em root}. A {\em forest}
contains multiple trees, which may be interlinked.
\begin{figure}[h]
\begin{centering}\includegraphics[width=0.4\columnwidth]{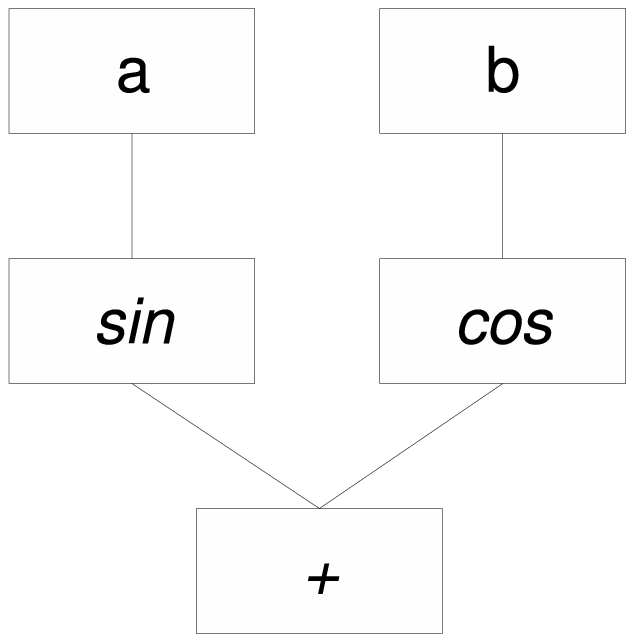}\par\end{centering}
\caption{\label{fig:simpletree}An expression tree.}
\end{figure}

A traditional expression tree corresponding to the expression $\sin a +
\cos b$ is shown in Fig.~\ref{fig:simpletree}. This illustrates the main
concepts of a tree: $a$ and $b$ are {\em leaf} nodes (having no children),
representing atomic components of the expression such as constants or variables;
``sin'' and ``cos'' are unary function nodes (performing some mathematical
operation on their children), and ``+'' is a binary function node. To compute
the value of the expression, we start at the leaves and propagate the values through
their parents, performing the appropriate mathematical operations along the way,
until we get to the root node (the ``+'' node, in this case), the result of
which is the value of the expression. 

In this traditional view, the result of the computation at any one node is a
single value, the value of some expression. MeqTrees goes a step further by
making the results {\em functions}. A typical MeqTree implements a real- or
complex-valued function of $N$ real variables, $f(x^{(1)},...,x^{(n)})$. The most
common variables -- at least in radioastronomy -- are time $t$ and frequency
$\nu$. For simplicity, we'll just use $t$ and $\nu$ in all further examples,
with the understanding that everything we describe can be generalized to
$N$-dimensional variable space. MeqTrees calls time and frequency {\em axes of
variability}, or simply {\em axes}.\footnote{The current implementation allows
up to 16 arbitrarily-named axes, though this number may easily be increased as
necessary.} In the example above, if we imagine that $a$ is a function of time,
$a=a(t)$, and $b$ is a function of frequency, $b=b(\nu)$, then the result of the
tree becomes a function of time and frequency, $f(t,\nu) = \sin{a(t)}+\cos{b(\nu)}$. 

MeqTrees represents functions as samples on a grid. To do this, we pick a {\em domain} in $t,\nu$, and define a gridding over that domain -- essentially, two vectors $(t_1,...,t_n)$ and
$(\nu_1,...,\nu_m)$. The function $f$ can then be represented by a
two-dimensional array of samples, $\lbrace f_{ij}=f(t_i,\nu_j)
\rbrace$. The result of the root node above -- the function $f(t,\nu)$
-- is then the array $\lbrace f_{ij}\rbrace$.

\begin{figure}[h]
\begin{centering}\includegraphics[width=0.8\columnwidth]{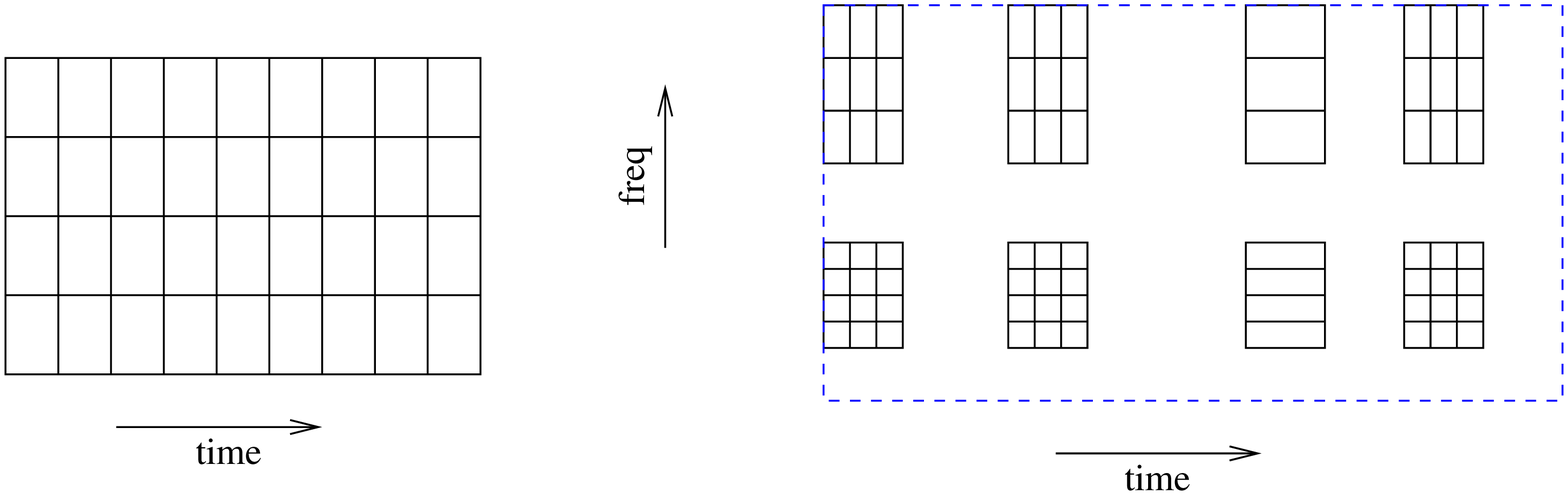}\par\end{centering}

\caption{\label{fig:domain}A node computes the value of some function on
a grid (which we call the {\em cells}) within some domain. The domain on the
left is two-dimensional ($t,\nu$) and has identical and contiguous cells on a
regular grid. It is also possible to have irregular grids of non-contiguous
cells with different sizes, as illustrated on the right (cell {\em size} is only
relevant for some rather obscure operations.) Higher-dimensioned domains are also
possible.}
\end{figure}

This then, in a nutshell, is the way MeqTrees work. A a {\em request} object is
created that contains (among other things) the two vectors, $(t_1,...,t_n)$ and
$(\nu_1,...,\nu_m)$. These are called {\em cells} in MeqTree parlance, since
they represent the rectangular cells of a two-dimensional grid. Note that the
grid stepping does not need to be regular -- see Fig.~\ref{fig:domain} for an
example. The request object is a request to compute a function (whatever
function happens to be defined by this tree) on a certain grid. This request is
passed to a node of the tree (usually the root node). To compute the function, a
typical node will pass the request on to its children, and then perform some
mathematical operation on the returned results. 

This also describes the interface to a node -- a node is given a request
(representing a gridding), and returns a result (representing a function sampled
on that grid). Indeed, since the interface to any tree is via its root node,
operationally a tree is indistinguishable from the root node. When parent nodes
deal with child nodes, they have no knowledge (nor do they need any) of whether
those child nodes are individual leaf nodes, or have whole subtrees hiding
behind them.

The result of a parent node is (usually) determined by performing some
mathematical operation on the results of its children. The exact operation is
determined by a node's {\em class}. Let's say a parent of class $H$ implements
the binary operation (or function) $H(a,b)$. If the results of its two child
nodes (or, equivalently, the subtrees rooted at their nodes) correspond to the
functions $f(t,\nu)$ and $g(t,\nu)$, then the result of the parent corresponds
to $h(t,\nu) = H(f(t,\nu),g(t,\nu))$. 

A leaf node has no children, and can compute its result (i.e. the function $f$
that it implements) in a self-contained way. This is also determined by its
class, for example:

\begin{description}
\item[MeqConstant] nodes return a constant value, $f(t,\nu)\equiv c$.
\item[MeqFreq] nodes return the frequency, $f(t,\nu)=\nu$.
\item[MeqParm] nodes compute, e.g., a polynomial
$f(t,\nu)=c_{00}+c_{10}\nu+c_{01}t$, where $c_{ij}$ are read from an external
parameter table (these are typically solvable parameters, as will be described
below.)
\item[MeqFITSImage] nodes return, e.g., an image of the sky read from a FITS
file. In mathematical terms, an image is a sampled brightness distribution,
$I(\nu,l,m)$. The sky coordinates $l,m$ are another example of axes of
variability.
\item[MeqSpigot] nodes interface to a Measurement Set (MS) used to store
observational data in radioastronomy, and return visibilities sampled by a
particular interferometer, $V(t,\nu)$. 
\end{description}

To sumarize, a typical tree computes a function defined on a grid in
$N$-dimensional variable space. A forest of trees computes a
collection of functions, which together constitute a numerical
model. The model may contain solvable parameters which may be
optimized in various interesting ways (Sect.~\ref{sec:solving}).

\subsection{Implementation\label{sec:Implementation}}

The implementation of MeqTrees consists of three main components:
  
\begin{enumerate}

\item The Tree Definition Language (TDL) is a Python-based language for building expression trees.
It allows one to succinctly specify nodes and their connections by means of class, name, children, 
and other options. 

\item The {\em meqserver\/} is the computational back-end of MeqTrees. It is mostly implemented in C++. 
A Meqserver process takes care of constructing and evaluating trees, and interfacing them to datasets.

\item The {\em meqbrowser\/} is a separate GUI (implemented in Python, and running in a separate process) for controlling meqservers. It parses TDL scripts, instructs servers to build the corresponding trees, and takes care of visualizing the results. (Note that it is also possible to run meqservers in non-interactive mode, without a browser.)

\end{enumerate}

\subsubsection{Specifying trees in TDL}

Fig.~\ref{fig:MeqBrowser} gives a (very simple) example of how to specify a tree using TDL. Part of a TDL script
is displayed in the middle section of the meqbrowser (see Sect.~\ref{sec:The-MeqBrowser-GUI}). TDL \citep{TDL-companion} is basically regular Python, plus some operator overloading that allows one to succinctly specify nodes and their connections. The TDL code shown in Fig.~\ref{fig:MeqBrowser} demonstrates the basic syntax for specifying nodes. 

Rather than always building trees from individual nodes, it is often far more efficient to manipulate higher-level Python objects and frameworks which, while presenting a simplified interface, cause trees
to be constructed behind the scene. This can hide a lot of unnecessary detail from the non-expert user, and accelerate development of complicated trees. Python as a language is very well-suited for
developing hierarchical object-oriented frameworks. We have developed a number of such frameworks in the context of radioastronomy: a lower-level framework called ``Meow'' (Measurement Equation Object
frameWork), and two higher-level frameworks for simulation
(``Siamese'') and calibration (``Calico''). More will surely follow.

\subsubsection{Meqbrowser and meqserver}

Being a semi-interpreted language, Python offers wonderful flexibility and ease of programming, but computing efficiency is not one of its stronger points. The actual computations in MeqTrees are performed by a fast, optimized back-end called the Meqserver, which is written mostly in C++. A GUI front-end called the meqbrowser (written in Python) provides a rich interface to the computational back-end. In a typical session, operations are divided as follows:
  
\begin{itemize}
\item The user loads a TDL script into the front-end (meqbrowser). This script
-- along with any associated option settings -- specifies the exact structure of
a tree (essentially, the structure of a computation).
\item Meqbrowser executes the TDL code, which results in a string of
instructions on how to assemble the tree to be sent to the back-end
(meqserver).
\item Meqserver constructs its internal tree representation based on the
supplied instructions.
\item The user operates the GUI to specify external data (e.g. a Measurement Set
to be calibrated, etc.); references to this data (pathnames, etc.) are also
passed to the meqserver.
\item The user operates the GUI to instruct the meqserver to start processing
data.
\item Meqbrowser monitors progress and (optionally, upon the user's instruction)
fetches and visualizes intermediate results (see
Sect.~\ref{sec:Visualization}).
\item It is also possible to bypass the GUI front-end, and operate the meqserver
noninteractively (i.e. in batch mode.)
\end{itemize}

Such an architecture allows for great flexibility in specifying how
computations are to be carried out (since all the specification is
done in TDL/Python), yet avoids the computational inefficiency
associated with scripting languages. A lot of thought has been put
into making the computational engine as efficient as possible (see
Sect.~\ref{sec:performance}). And while in principle a MeqTrees-based
computation cannot match the theoretical performance of hand-optimized
compiled code, in real-life testing its performance has proven to be
either equivalent, or (in the worst case) within a factor of 2--3 of
hand-optimized implementations.

\subsubsection{\label{sec:The-MeqBrowser-GUI}The meqbrowser GUI}

\begin{figure}
\begin{centering}\includegraphics[width=0.95\columnwidth,keepaspectratio]{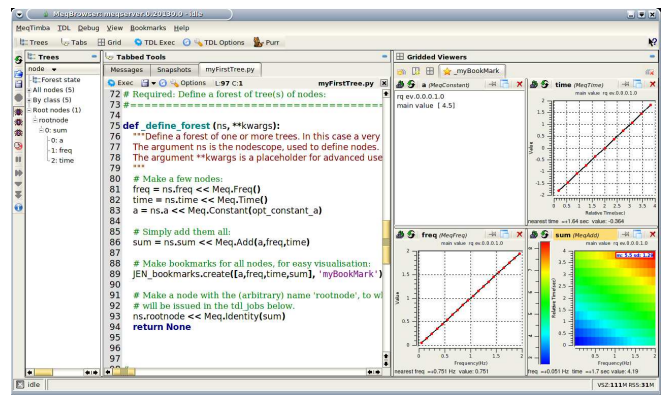}\par\end{centering}
\caption{{\em \label{fig:MeqBrowser}The meqbrowser GUI.}}
\end{figure}

The meqbrowser GUI is composed of three main sections, and a choice of menus and other buttons (Fig.~\ref{fig:MeqBrowser}). In this example, the contents of the TDL
script {\tt myFirstTree.py} has been loaded by means of the TDL menu
at the top, and is displayed in the middle section. The TDL code in
the {\tt \_define\_forest()} function defines the nodes that make up
this very simple tree. The Python code may be edited directly in the
browser, or with an external editor. After compilation (i.e.
generation of nodes on the meqserver side), the tree is displayed in the
left section, where it may be browsed by opening and closing
branches. Clicking on a node displays its contents in various ways in
a panel in the right section, which may then be inspected
interactively. The tree is executed by issuing a request to a named
node (here called 'rootnode'). The execution is done by means of a
{\tt \_tdl\_job()} function in {\tt myFirstTree.py}, which may be
called via the {\tt TDL Exec} button. Upon execution, the various
display panels on the right will come alive, showing the node results.

A TDL script may have compile-time options, whose values may be set
interactively in a popup window under the {\tt TDL Options} button.
The script may also have run-time options, which may be set in a popup
window under the {\tt TDL Exec} button. It also offers a choice of
available ``TDL jobs'' to be executed. Clicking on one of them
executes the forest, or performs some other function like invoking the
CASA imager. Currently selected values of TDL options are retained
in a configuration file, which greatly increases the ease of using
MeqTrees in practice. The configuration file also comes in handly when
a TDL script needs to be executed in batch mode, without the
meqbrowser.

Obviously, we have described only the basic functionality of the
meqbrowser, which has many powerful features to make it easy (and
fun!) for the user to execute trees and to inspect node results. For
instance, it has a bookmark button that offers shortcuts to the
display of predefined groups of nodes. The Purr button (``Purr is
Useful for Remembering Reductions'') activates a rather useful scheme
to save and describe all intermediate steps of a data reduction
project. Debugging functionality (stop, step, resume etc) is available
along the left edge. It also has a profiler, which measures the
processing of all nodes, either individually or by class. Execution
progress messages are displayed along the botton, and any errors may
be inspected in detail.

\subsubsection{\label{sec:Visualization}Visualization}

One of the cornerstones of MeqTrees is its emphasis on visualization
at all levels. This is based on the conviction that the quickest way
to develop and debug an algorithm is to be able to see what is going
on. It cannot be stressed enough that all this visualization is {\em
optional}, and does not affect computational efficiency when it is not
used. We expect that this capability will be very popular, and will
soon become the norm. Therefore we envisage a steady increase in
sophistication of both standard and application-specific
visualization.

In addition to making visualization possible, the meqbrowser also
makes it easy. Clicking on any node in the meqbrowser will display
either the node status record (very instructive!), or a specific
field, or a plot of the latest result in its cache. Various different
types of plots or representations may be selected by right-clicking on
a node. When displaying images, middle-clicking on the plot itself
will produce a vertical and horizontal cross-section through that
point. Optionally, flags may be indicated in the plots. As illustrated
in Fig.~\ref{fig:MeqBrowser}, the plotter will adapt automatically to
the dimensions of the displayed result: frequency, time, both, etc. If the
result has more than 2 dimensions, different cross-sections may be
selected. There is also a multidimensional plotter.

Some nodes display their results in specific ways. The MeqSolver node
produces plots that indicate the quality of the solution, and its
evolution over successive iterations. The ultimate goal is some kind
of visualization of the $\chi^{2}$ surface. The MeqComposer node
produces a so-called {\em inspector} plot, i.e. averaged 
time-tracks of its multiple results, side-by-side in the same panel.  
The inspector is a cumulative plot in the sense that
the time-tracks just grow in length with successive results. The MeqDataCollect
node also displays the results of multiple nodes in the same plot,
but is refreshed each time. It offers two modes, spectra or real vs. imaginary,
and is hierarchical: the results of various MeqDataCollect nodes may
be combined in the same plot, e.g. with different colors or symbols.

Ease of use is greatly increased by a menu of meqbrowser {\em bookmarks}.
Clicking on a bookmark conjures up the visualization of a specific
node, or a {\em bookpage} of associated nodes. Bookmarks are defined
by the tree designer, to highlight particular aspects of what the
tree is doing. Selected bookpages remain close at hand by means of
tabs (see Fig.~\ref{fig:MeqBrowser}). Nodes make their information
available for display by {\em publishing} it, i.e. sending it off
into the void, to be picked up by another program. A node may be induced
to publish every time it gets a new result, which is the default for
bookmarked nodes. This makes it easy to watch intermediate
results while the tree is executing a sequence of requests.

Although the standard visualization of MeqTrees offers substantial
functionality, we expect that many users will develop specialised
visualisation nodes for their specific application areas. This will
be encouraged and supported, for instance by offering nodes and other
tools (like a result object) to make this easier. For the moment,
the user-definable nodes PyNode and PrivateFunction should do. 

\subsubsection{On the choice of languages}

While for a lot of scientific software, programming language is chosen on the basis of nothing more than
the author(s) personal preference and/or proficiency, and needs no further justification, MeqTrees development has taken the unusual route of combining two languages (C++ and Python). The reasons for this perhaps require further elaboration.

Python is sometimes thought of as ``only'' a scripting language, but in fact it is a powerful high-level language for structured and object-oriented programming. A number of very capable Python libraries and frameworks for scientific programming have emerged in recent years (most prominently numpy/scipy\footnote{{\tt http://www.scipy.org}}), and it also offers bindings to C++ GUI and visualization toolkits such as Qt and Qwt. This makes it feasible to write large and feature-rich applications completely in Python (the meqbrowser being a case in point). Casual programmers find it easier to pick up than a compiled and strongly-typed language such as C++, which is why many astronomers dabble in Python scripts, but very few dare to write C++ programs. On the other hand, programmers equally proficient in both can usually (in the authors' personal experience, at least) implement a given algorithm in Python much more rapidly than in C++. These are the reasons behind Python's recent success and adoption as the high-level language component of many projects, including CASA and MeqTrees (specifically, TDL.)

The question then remains, why use C++ at all. Being semi-interpreted and late-binding, Python has two drawbacks. The first of these is that errors (with the exception of syntax) are only detected when the offending piece of code is actually executed, whereas a compiled language will catch many errors at the compilation stage. This can lead to some unusual bugs, and makes proper test coverage all the more important, especially for large applications. In the authors' experience, however, this consideration is far outweighed by the considerably increased development speed. It is probably true that a poorly-tested Python application will tend to have more bugs than a poorly-tested C++ application (though the two will probably be equally useless!) On the other hand, the Python application will have been developed much faster, leaving much more time to find and fix the bugs. This drawback, therefore, only really applies to 
poorly-tested code.

The second -- and far more fundamental -- drawback of Python is that a semi-interpreted language is inherently slower to execute than compiled code. Any computation requiring a large number of iterations over multiple lines of code (such as e.g. the nested loops so often appearing in numerical programming) becomes grossly inefficient when implemented in Python. In addition, the relatively high abstraction level of the language makes it practically impossible to optimize for memory access and CPU cache usage, which makes it a poor choice for High-Performance Computing (HPC) applications. These remain the domain of low-level languages such as FORTRAN, C and C++.

A common way to get the best of both worlds (easy and rapid development, plus computational efficiency) is to implement the core computations in a low-level language, while using Python as a high-level binding. Different projects take a different approach to where they put the language boundary. For example, CASA can be thought of as coarse-grained, since it consists of rather large tasks written in C++, with Python used for high-level task control and ``glue''. At the other end of the spectrum, the numpy/scipy library is very fine-grained, implementing array operations (including some rather advanced operations such as FFT, filtering, statistics and morphology) in C, while leaving the rest of the algorithm to be implemented in Python. MeqTrees occupies a position somewhat between the two (while in fact adopting some components of both.)

Note that the Python component of MeqTrees is implemented as a layer on top of the core C++ libraries. The tree-building and evaluation layer has no awareness of nor dependence on Python. In principle, this means that the core libraries can be directly called as a C++-only toolkit. The success of TDL, however, has meant that we have not (to date) explored this possibility.

\section{Data Model}

We have described how MeqTrees uses trees to represent mathematical
expressions that comprise a numerical model. This is only half of the
story; the other half is the actual computation, i.e. what kind of
data can be fed into these expressions, and how efficiently can it be
processed. The capabilities of MeqTrees are in large part determined
(and limited) by this underlying data model. This section
describes the data model in more details.

\subsection{\label{sec:grids}Grids and functions}

\begin{figure}[h]
\begin{centering}\includegraphics[width=0.4\columnwidth]{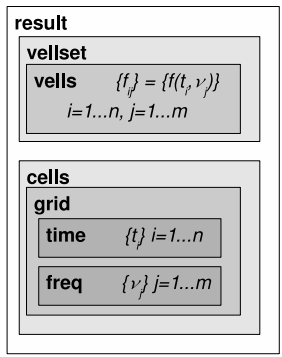}\par\end{centering}
\caption{\label{fig:result}The layout of a basic result object.}
\end{figure}

As stated previously, the atomic unit of computation in MeqTrees is a
{\em function} represented by a set of samples over an $N$-dimensional
grid, e.g., a function of frequency and time $f(t,\nu)$. Internally,
this is represented by a {\em cells} object specifying the grid --
containing vectors of, e.g., times $(t_i)$ and frequencies $(\nu_j)$
-- and an $N$-dimensional array of samples,
e.g. $(f_{ij}=f(t_i,\nu_j))$. For historical reasons the latter object
is called a {\em vells}. A vells object is placed into a container
called a {\em vellset} (the rationale for this will be explained in
Sect.~\ref{sec:solving}.) A cells and a vellset together then
constitute a {\em result} object (Fig.~\ref{fig:result}). A result
object is the unit of data that is passed between nodes.

If a function does not vary over a given axis, then the dimension of
its vells along that axis can be equal to 1, i.e. it can be represented
by fewer actual data points. For example, given a cells of $M$ times
and $N$ frequencies (and assuming time is the first axis and frequency
is the second axis in our grid -- the ordering of axes is fixed prior
to computation), a function can be represented by vells with
dimensions of:
   
\begin{description}
\item[$M\times N,$] for a function variable both in time and frequency;
\item[$M\times 1,$] (or, equivalently, just an $M$-vector) for a function variable only in time;
\item[$1\times N,$] (this is {\bf not} equivalent to an $N$-vector, since frequency is the {\bf second} axis);
for a function variable only in frequency;
\item[$1,$] for a constant value.
\end{description}

All mathematical operations transparently handle vells of different
dimensions. For example, if the "+" node in Fig.~\ref{fig:simpletree}
receives an $M\times 1$ array from one child and a $M\times 1$ array
from the other child, it will perform $M$ additions and return a
$M\times 1$ array (i.e. a function of time only). If, on the other
hand, it receives an $M\times 1$ array (function of time) and a
$1\times N$ array (function of frequency), then it will perform
$M\times N$ operations and the result will be an $M\times N$ array
(function of time and frequency). These decisions are made directly in
the tree at runtime, so the exact same tree can be used to compute an
expression involving only constants, or an expression involving
functions of time, frequency, etc. In the latter case the computation
is automatically optimized in the sense that the tree keeps track of
what values depend on what axes, and only executes the mimimum number
of calculations.

This is actually one of the most powerful features of MeqTrees. It is
often the case in numerical modeling that one starts with a simple model (i.e. constant parameters), and adds complexity (e.g. parameters with time dependence) later. With traditional code, adding a time
dependency to a particular branch of a calculation requires that arrays be resized, for-loops added to iterate over time, bugs introduced and squashed again, etc. Depending on the complexity of a piece of code, this can become quite a daunting task, if not an insurmountable barrier to experimentation. With MeqTrees, you only
need to change the property of a parameter up in the tree, and time dependence is then automatically propagated throughout all calculations.

\subsection{\label{sec:tensors}Scalars and tensors}

Scalar functions have a single real or complex value at each grid
point. A tensor function of type (dimensionality) $n_1,n_2,...,n_k$
has $n_1\times...\times n_k$ values for every grid point. For example,
the 2$\times$2 coherency and Jones matrices used in the RIME are
type-2,2 tensor functions. An efficient representation of tensor
functions is therefore important for efficient implementations of the
RIME.

\begin{figure}[h]
\begin{centering}\includegraphics[width=0.8\columnwidth]{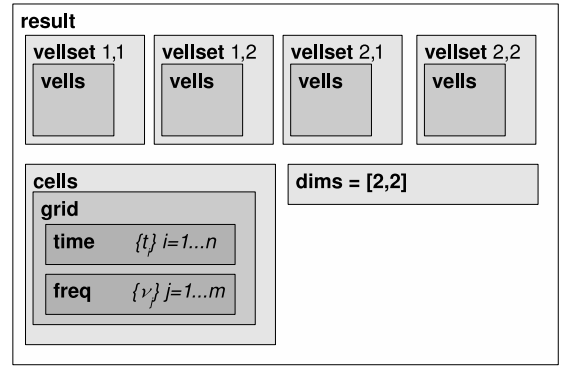}\par\end{centering}
\caption{\label{fig:result-tensor}The layout of a tensor result object.}
\end{figure}

MeqTrees represents tensors by a result object containing a list of
$n_1\times...\times n_k$ vellsets, and a vector of integers
$(n_1,...,n_k)$ describing the
dimensionality. Figure~\ref{fig:result-tensor} shows a tensor result
corresponding to a $2\times2$ matrix. Each matrix element is
represented by a separate vellset (though all share a common cells
object, thus a common grid definition.) This means, among other
things, that each element of a matrix can have independent axes of
variability. Operationally, it is quite common to see diagonal
matrices where the diagonal elements are functions of frequency and/or
time, while the off-diagonal elements are zero (and thus constant.) 
Such tensors can be most efficiently represented with this scheme.

Most node classes can transparently accept tensor arguments. The
normal convention is to perform the corresponding mathematical
operation element-by-element. All arguments must then have the same
tensor dimensions (or, as a special case, be scalar, in which case the
scalar value is reused for all elements), and an error is reported
otherwise.

MeqTrees also provides a few specialized tensor nodes. The
MatrixMultiply node is used to multiply matrices (as well as
vectors). The MatrixInvert22 node inverts 2$\times$2
matrices.\footnote{General matrix inversion is not yet implemented,
since 2$\times$2 matrices are sufficient for RIME purposes. It could be
added as needed.} The Composer can combine the results of multiple
nodes into a single tensor, and the Selector extracts individual
tensor elements.

\section{\label{sec:flags}Data flags}

Radio astronomy has to operate in an environment in which there are
many external and internal sources of Radio Frequency Interference
(RFI). It is therefore important to be able to flag bad data values,
and propagate these flags throughout all calculations, so that results
derived from bad data are also properly flagged and ignored as
necessary.

\begin{figure}[h]
\begin{centering}\includegraphics[width=0.8\columnwidth]{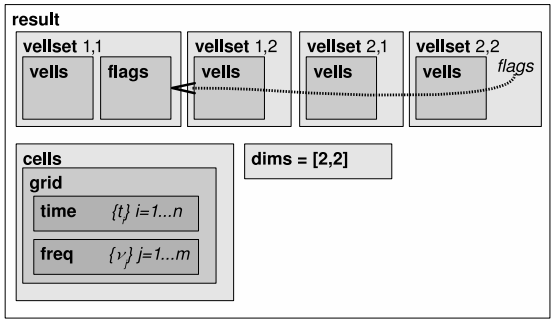}\par\end{centering}
\caption{\label{fig:result-flags}The layout of a result object with flags. Note that vellsets 1,1 and 2,2 share the same flags, while vellsets 1,2 and 2,1 have no flags at all.}
\end{figure}

MeqTrees can associate a {\em flag vells} with each vellset of a
result. A flag vells is an array of integer flagwords that follows the
same dimensionality rules as for normal vells arrays (see
Sect.~\ref{sec:grids}), i.e. given a cells of $N$ time and $M$
frequency points, a flag vells may have dimensions of 1, N, $1\times
M$, or $N\times M$. The latter case associates a separate flagword
with every time/frequency point -- a raised bit in the flagword
$w_{ij}$ indicates bad data at $t_i,\nu_j$. The intermediate cases
correspond to entire times or frequencies being flagged at once, while
the first (and admittedly not very useful) case has a single flagword
for the entire domain. The different vellsets of a tensor result may
have different flag vells, or may share flags by reference (see
Fig.~\ref{fig:result-flags}).

Flag vells are automatically propagated through tree nodes in a
mathematically sensible manner. For example, when an Add node receives
flagged results from its children, the flag vells associated with its
result is a bitwise-OR of all the child flags. This means that at the
root of the tree, the final result will have flags for every $t,\nu$
point where the resulting value is derived from something that was
flagged. Statistical and reduction operations (such as taking the mean
over a certain set of axes) will also ignore flagged values when
computing the statistics.

The flagword contains 32 individual bits, which allows for some very
versatile flag management. Since flags can be built up via different
procedures (automatic flagging algorithms, heuristics based on
metadata describing system status during measurement, manually raised
flags), it is exteremely useful to associate these different {\em
flagsets} with different bit positions in the flagword. The user then
has the option of activating or ignoring specific flagsets via a
bitmask of currently active flags.

Flags can be preserved under storage, if the data format supports
them. For radio astronomical data, a standard storage type is the
Measurement Set \citep{MS2definition}, which defines standard columns
for boolean flags. MeqTrees extends this by specifying an additional
column of bitwise flags.

\subsection{\label{sec:flagging}Flagging}

Flags can also be generated directly inside a tree. MeqTrees offers a
very simple but versatile scheme for this. The MeqZeroFlagger node
sets flags in its child result based on a comparison to zero. The user
then has to supply some subtree that produces a result that can be
used as a discriminator, e.g. in which ``bad'' data corresponds to
values greater than zero. The MeqMergeFlags node is then used to merge
the new flags with those of the original data.

Since the discriminator expression is supplied by an arbitrary
subtree, any type of flagging can be implemented, from simple data
clipping, to flagging based on the value of some completely unrelated
expression defined over the same domain (such as the value of a
solution, see Sect.~\ref{sec:solving}).

\section{\label{sec:solving}Parameters and model fitting}

Consider a mathematical model, i.e. some function $M(\nu,t)$ that depends on
a number of parameters $(p_1,...,p_k)=\vec p$. We denote this as
$M(\nu,t;\vec p)$. {\em Model fitting} is the process of finding a
value of $\vec p$ that minimizes the difference (according to some 
predefined metric) between measured data $D_{ij}=D(\nu_i,t_j)$ and $M_{ij}(\vec
p)=M(\nu_i,t_j;\vec p)$. We also call this {\em solving} for $\vec
p$. In radio astronomy, the model $M$ is given by some
parameterization of the RIME, and the process of model fitting is
called {\em calibration}. This is explained in more detail in Sect.~\ref{sec:RIME},
here we first want to describe the general approach to model fitting employed in MeqTrees.

\subsection{Solving in MeqTrees}

\begin{figure}[h]
\begin{centering}\includegraphics[width=0.8\columnwidth]{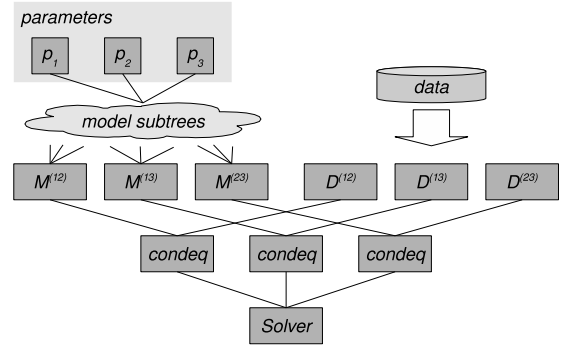}\par\end{centering}
\caption{\label{fig:solver}A schematic layout of a solver tree.}
\end{figure}

The MeqTrees approach to solving is as follows (Fig.~\ref{fig:solver}). 
A set of subtrees of arbitrary complexity implements the model $M$.  
The model is not necessarily a single
function, but can be a whole set of functions, e.g. $\lbrace M^{(pq)}
\rbrace$, giving the visibilities per baseline $p-q$ (see Sect.~\ref{sec:RIME}). 
These model trees can be quite similar to the simulation trees discussed in that section.
At certain parts of the model are the unknown parameters -- these are represented by MeqParm nodes. MeqParms are initialized with best guesses (or previous solutions read from
disk, if available.)

A parallel set of subtrees provides the data $D^{(pq)}$. These
subtrees can be (and usually are) as simple as a single MeqSpigot node
(which interfaces to a Measurement Set containing visibility data), but 
can also contain preprocessing steps as needed. The two sets of subtrees are linked up
via MeqCondeq nodes, which are in turn all children of a MeqSolver.

At the beginning of a solution, the MeqSolver issues a special request
that designates a subset of the MeqParms as solvable. Subsequently,
trees that contain solvable MeqParms, when asked to compute a result,
automatically augument it with partial derivatives with respect to the solvable
parameters $\partial M^{(pq)}/\partial p_k$. Note that this is
completely transparent to the user -- any tree that can compute a
function can also compute the derivatives of that function (see
below).

MeqCondeq nodes then form up the difference
$\Delta^{(pq)}=M^{(pq)}-D^{(pq)}$, and also the partial derivatives
w.r.t. each solvable parameter, $\partial \Delta^{(pq)}/\partial p_k$,
and return these to the solver. The solver uses some algorithm to
determine a set of incremental parameter updates $\Delta \vec p$, and sends
these updates back up the tree to the MeqParms, at which point the
procedure is repeated until convergence has been achieved, or a
maximum number of iterations has been reached.

\subsection{Estimating the derivatives}

\begin{figure}[h]
\begin{centering}\includegraphics[width=0.8\columnwidth]{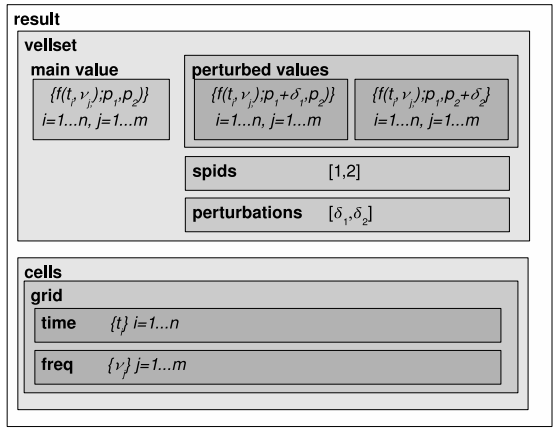}\par\end{centering}
\caption{\label{fig:result-spids}The layout of a result object with perturbed values.}
\end{figure}

MeqTrees currently estimates first derivatives via finite
differencing. If a subtree implementing the function $f(\nu,t)$
depends on the solvable parameters $p_1$ and $p_2$, then the vellset
in its result (Fig.~\ref{fig:result-spids}) will actually contain
three vells: the ``main'' value $f(\nu,t;p_1,p_2)$, plus two
``perturbed'' values $f(\nu,t;p_1+\delta_1,p_2)$,
$f(\nu,t;p_1,p_2+\delta_2)$.

If another subtree for $g(\nu,t)$ depends on solvable parameters $p_2$
and $p_3$, then its result will likewise contain three values:
$g(\nu,t;p_2,p_3)$, $g(\nu,t;p_2+\delta_2,p_3)$,
$g(\nu,t;p_2,p_3+\delta_3)$. Note that each vellset also contains a
vector of {\em spids} (solvable parameter identifiers), which indicate
what solvable a particular perturbed value is associated with.

Now consider how this information propagates down the tree. If $f$ and
$g$ are the child nodes of a MeqAdd node returning
$h(\nu,t)=f(\nu,t)+g(\nu,t)$, then the addition must be executed four
times:

\begin{eqnarray*}
h(\nu,t;p_1,p_2,p_3) & = & f(\nu,t;p_1,p_2)+g(\nu,t;p_2,p_3) \\
h(\nu,t;p_1+\delta_1,p_2,p_3) & = & f(\nu,t;p_1+\delta_1,p_2)+g(\nu,t;p_2,p_3) \\
h(\nu,t;p_1,p_2+\delta_2,p_3) & = & f(\nu,t;p_1,p_2+\delta_2)+g(\nu,t;p_2+\delta_2,p_3) \\
h(\nu,t;p_1,p_2,p_3+\delta_3) & = & f(\nu,t;p_1,p_2)+g(\nu,t;p_2,p_3+\delta_3) \\
\end{eqnarray*}

Note that the left-hand side contains the ``main'' value of $h$ plus
three perturbed values of $h$ with respect to $p_1,p_2,p_3$, while the
right-hand side contains a mix of main and perturbed values of $f$ and
$g$, combined according to a simple looping algorithm. This kind of
loop over perturbed values is automatically executed inside every
MeqTrees node, thus ensuring that the root of the tree computes
perturbed values for every solvable parameter in the tree.

Given a main and a perturbed value, the actual derivative is estimated as:

\[
\frac{\partial f}{\partial p_k}(\nu,t;p_k) \approx \frac{f(\nu,t;p_k+\delta_k) - f(\nu,t;p_k)}{\delta_k}
\]

Note that we keep writing $\nu,t$ here to emphasize the fact that both
$f$ and its derivatives are potentially functions of many variables
such as frequency and time. That is, a separate value and a separate
derivative are computed at every time-frequency point.

\subsection{Least-squares solver}

A (weighted) least-squares solver minimizes the difference $D-M$ in a
least-squares sense, i.e. finds a $\vec p$ that minimizes the
chi-squared sum:

\begin{equation}
\chi^2(\vec p) = \sum_{ij}w_{ij}^2(D_{ij}-M_{ij}(\vec p))^2
\label{eq:chisq}
\end{equation}

where $w_{ij}$ are (optional) weights. Different methods of minimizing
$\chi^2$ are known; the suitability of a particular method depends on
properties of $M$ such as degree of linearity with respect to $\vec p$,
etc. Since MeqTrees can implement models of arbitrary complexity, our
initial designs have tended towards a ``policy-free'' solving scheme
that works adequately in most cases but is not necessarily the optimal
one for each particular case. We have therefore decided to use the
AIPS++/CASA solver \citep{WNB:solver}, which implements the
Levenberg-Marquard algorithm \citep{Madsen:LM}. The LM algorithm is a
type of gradient descent method which is particularly well-suited to
nonlinear problems.

An important problem with any solver is the handling of
ill-conditioned problems (i.e. when there's not enough information to
solve for all unknowns.) The AIPS++/CASA solver works by accumulating
normal equations (received from the MeqCondeqs), then inverting the
solution matrix. The inversion is done by SVD, which detects
ill-conditioning and handles it by effectively reducing the number of
unknowns (this is called reducing the rank of the solution.) A number
of solver diagnostics, including rank and condition numbers, is
automatically generated by the MeqSolver node, and may be visualized
by the user to get an indication of the quality of the fit.

\subsection{Alternative approaches}

Other kinds of solvers are being actively considered for inclusion in
MeqTrees. These may require different ways of calculating the derivatives:

\begin{itemize}

\item Analytic derivatives are known to produce more stable solutions. 
MeqTree nodes can be adapted to compute and propagate analytic derivatives 
via the chain rule (and fall back to finite differencing should a node be 
encountered that cannot compute analytic derivatives). 

\item Second derivatives may allow for better solvers. Second derivatives 
may be computed via double-differencing, or analytically. 

\item Bayesian solvers, rather than using derivatives, sample the function over a ``cloud'' in parameter space. That is, they generate a random set of vector offsets $\vec\delta_1,...,\vec\delta_n$, and compute the perturbed model at each offset $M(\vec p+\vec\delta_k)$. Note that our current scheme of computing perturbed values with respect to each solvable parameter can be considered a special case of this, each vector offset $\vec\delta_k$ being a simple shift along axis $k$ in parameter space, orthogonal to all the other offsets.

\end{itemize}

In principle the MeqTree code and internal data structures can be
easily adapted to any of the approaches listed here.

\subsection{MeqParms}

A MeqParm node represents solvable parameters of the model. One of the
most powerful features of MeqTrees is that each parameter can be a
function of $\nu,t$ (and, naturally, any other dimensions.) The
current implementation provides a polynomial, so the actual solvables
are coefficients of the polynomial. The degree of the polynomial may
be specified separately for each MeqParm, and for each solution. Other
smooth functions of $\nu,t$ may of course be obtained by combining
polynomial MeqParms into subtrees.

MeqParms can store their solutions to {\em MEP tables}. The solutions
are identified by domain (e.g. in $\nu,t$). A typical calibration
procedure involves solving for one subset of parameters, storing these
solutions, then going on to solve for another subset, while using the
stored solutions of the first set when evaluating the model. MEP
tables have a Python interface, so the stored solutions can be
analyzed, plotted and/or reprocessed with external tools.

\subsection{\label{sec:subtiling}Continuity and solve domains}

In many problems, radio astronomy not excepted, data volumes preclude
processing an entire observation at once. Instead, data has to be broken up 
into chunks along the e.g. time and/or frequency axes, and processed 
sequentially. MeqTrees calls such chunks {\em tiles}. When solving for
parameters, solutions are by necessity generated on a tile-by-tile basis. On
the other hand, physical considerations can often provide continuity
constraints between tiles, and it is important to take advantage of these constraints.

In the simplest case, a MeqParm will generate one solution per tile,
and use that solution as the starting value for the next tile. No
explicit continuity constraint is imposed. In this case one makes the tile size 
small, in accordance to how quickly a parameter is expected to vary. It
is possible that the expected variation in a parameter is slower than
the largest practical tile size. An extreme case of this is when
trying to solve for a single value across the entire measurement. If
the data for the entire measurement does not fit in memory, obtaining
a global solution requires multiple runs through the data, which may 
impose unacceptable I/O penalties. In general this is a very thorny 
problem. One (rather inelegant) way around this is doing
tile-by-tile solutions with the largest possible tile, and smoothing
the solutions afterwards. Other options are averaging the data, or extracting a strided 
subset of the data. 

A more complicated case relates to scenarios where we want to
simultaneously solve for parameters that have different degrees of
variability. A typical example from radio astronomy would be receiver
phases (which are almost constant in frequency, but vary rapidly in
time) and bandpasses (which vary very slowly in time, but have very
complex behaviour in frequency). MeqParms address this via a technique
called {\em subtiling}. Each MeqParm may be setup with its own subtile
size, and an independent solution is then done within each subtile of
the larger overall tile. In the example here, phases would have a
subtile of size 1 in time, and bandpasses would have a subtile of size
1 in frequency. A separate phase solution per timeslot (constant
across all frequencies) and a separate bandpass per frequency
(constant across all times in the tile) would then be obtained.

The combination of tile size, subtiling and polynomial degrees makes
for a very flexible way to specify parameter behaviour. It is in fact
a challenge to present all these options to the user in a non-bewildering 
fashion.

\section{\label{sec:performance}Performance considerations}

Given the large data volumes produced by radio astronomical
instruments (and the even bigger volumes required to simulate
instruments of the future such as the SKA), computational performance
is always going to be an important issue, both in terms of processing
speed and memory use.

With any software, there is generally a tradeoff between flexibility
and efficiency. Highly optimized code is by its nature difficult to
revise and extend, and vice versa. MeqTrees tries to get around this
problem by providing highly optimized building blocks (i.e. nodes),
while offering maximum flexibility in putting them together.

There is always some overhead associated with navigating a tree and
passing results around, but this overhead is independent of domain
size, whereas actual computational cost increases linearly with the
number of, e.g., time and frequency points. This implies that MeqTrees
is at its least efficient when using single-cell domains, where
housekeeping overhead dominates, and at its most efficient when using
large domains, where computational cost dominates. In practice, when
using domains of 500--1000 cells, the performance of MeqTrees becomes
comparable with that of hand-optimized code. This is achieved through
a number of generic optimization techniques, which will be described
below.

\subsection{Optimal use of axes}

Recall from Sect.~\ref{sec:grids} that a vells represents the value of
a function using the minimum required number of axes of
variability. Values with only a time or only a frequency dependence
are passed around as vectors, and are expanded to arrays only when
both a time and frequency dependence arises. This avoids redundant
computation. It is also possible to explicitly structure trees to take
advantage of this, i.e. to introduce extra axes as ``late'' in the
computation as possible.

\subsection{Result caching}

Each node maintains an optional result cache, which allows
computations to be reused. A straightforward but very powerful scheme
of {\em dependency tracking} allows a node to figure out exactly when
a result may be usefully cached. For example:

\begin{itemize}
\item A node with multiple parents should cache its result until all parents have retrieved it.
\item When solving, a result with no dependence on solvable parameters can be cached until the next iteration. Results that do depend on solvable parameters are never cached, since they're updated with each iteration.
\item A result with no dependence on time can be cached until the next tile (assuming we're iterating over time.)
\item If all parents have cached their results, a child may discard cache.
\end{itemize}

In practice, this means that only the minimum necessary part of the
tree is reevaluated when going from one solver iteration to the next,
or from one tile to the next. It is also possible to fine-tune the
caching policies to trade off computing time vs. memory footprint,
etc.

\subsection{Parallelism}

In these days of cheap computing, parallel processing is the obvious
approach to large computational problems. MeqTrees has been designed
and implemented with this in mind. The tree paradigm provides
ample opportunities for parallelisation, since different branches of
the tree (and different trees of the forest) may be executed
concurrently. On the other hand, the fact that trees may (and usually
do) share branches towards the top makes for interesting scheduling
problems -- it's not much use to execute branches in parallel if
they are going to spend most of their time waiting for the result of a
single shared sub-branch.

Meqserver has supported multithreading for a long time, so
multiple-core machines may execute different branches of a tree
concurrently. It employs a worker thread pool scheme, which avoids
shared-branch bottlenecks. If a thread becomes stuck waiting for the
result of a shared branch, another worker thread is woken up and
assigned to a different part of the tree. In practice, this means that
on a modern four-core machine, MeqTrees will happily keep all four
cores fully occupied (as long as there's sufficient paralellism in the
tree itself.)

Parallelism across a cluster is a far trickier proposition -- one has
to consider not only scheduling problems, but also cost of data
transport between cluster nodes. In 2008, the first MPI version of
MeqTrees was tested on a cluster in Oxford, with an eye on large-scale
simulations for the SKA radio telescope.\footnote{For the time being, it will
only be made available to users with special capabilities.} This
version allows parts of the tree to be distributed across a cluster,
and uses MPI to pass results between children and parents that reside
on different nodes. This version was tested across 8 cluster nodes and
scaled reasonably well, albeit on a rather simple simulation problem.
It is clear that the biggest amount of thinking needs to be put into the 
problem of how to distribute a given tree across a cluster
efficiently.

\section{\label{sec:RIME}The Measurement Equation of a Generic Radio Interferometer (RIME)}

The pressing need for 3GC is uncontroversial, but its specific development may follow various routes.
Important issues are the dynamic range limitations caused by bright sources, the application of DDEs, and the imaging and deconvolution of residuals. We propose to expose our approach to the DDE problem in a follow-up paper, and avoid discussing specific calibration schemes here. Some examples, however, will be necessary, in order to show how MeqTrees can support such flexible schemes (and, indeed, why it was designed the way it was.)

What is clear is that the Measurement Equation (RIME) has a vital role to play. In this section, we will therefore discuss the general structure of the RIME, and some of its properties. We hope to make clear how elegant it is, and
thereby to smooth the path to its wider adoption by the radio astronomical community. We will also indicate how the universality and modularity of the RIME opens the way for MeqTrees to offer a set of generic TDL processing scripts, which may be quickly adapted to a wide range of experiments with any radio telescope.  The latter is done by means of the simple expedient of plugging in different Jones matrices, or a different sky model. Note that the description of the RIME given here is by necessity qualitative and brief, since we will concentrate on how the RIME pertains to MeqTrees rather than give a full formal exposition.

The RIME was formulated by \citet{ME1} following preparatory work by \citet{Morris:1964}. \citet{ME4} then rewrote it in $2\times2$ matrix form, which is the one we follow here.\footnote{Some versions of the RIME still use $4\times4$ Mueller matrices. This is entirely equivalent, but much less transparent for our purposes.} Note that, since all the existing 2GC packages were written before the RIME was formulated, they {\em implicitly} implement a limited and approximate form of the RIME, usually optimized for a specific telescope.

The RIME is a {\em formalism} rather than one specific equation, and so it may be written down in many forms. 
A particularly elegant and simple form describes a ``mostly empty'' sky of discrete sources. In this form of the RIME, the predicted value of the visibility sample $\vec V_{pq}$ is given by:

\begin{equation}
\vec V_{pq} = \vec G_{p} \left( \sum^{N}_{k=1} \vec E_{pk} \vec X_{k} \vec E^{\dagger}_{qk} \right) \vec G^{\dagger}_{q}
\label{eq:rime}
\end{equation}

where $\vec V_{pq}$ is the $2\times2$ visibility (also called {\em coherency}, or {\em uv-data})
matrix measured by the interferometer formed by stations $p$ and $q$. The sum is taken over the contributions $\vec X_{k}$ from $N$ discrete sources in the field, at positions $l_{k},m_{k}$. They are
corrupted by instrumental effects that are represented by so-called Jones matrices \citep{Jones}. All the terms
of eq.~(\ref{eq:rime}) are $2\times2$ matrices, and ``$\dagger$'' represents the Hermitian (or conjugate) transpose operator. The $\vec E_{pk}$ term is itself product of a number of Jones matrices corresponding to {\em direction-dependent effects (DDEs)} associated with station $p$ and direction $l_{k},m_{k}$, while $\vec G_p$ is a product of Jones matrices for the {\em direction-independent effects (DIEs)} associated with station $p$. 

Note that eq.~(\ref{eq:rime}) assumes that all instrumental effects are station-based, i.e. can be fully described by Jones matrices, each associated with a particular station $p$. This is called the {\em selfcal assumption}. It is crucial because it increases the ratio between the number of equations (given by measured uv-data) and independent unknowns to the level where selfcal generates non-trivial solutions. In principle the observed data is also corrupted by {\em interferometer-based} errors (also called closure errors), which are conventionally modelled via extra multiplicative and additive terms on the elements of $\vec V_{pq}$. These can then be solved for (with some care), assuming the errors are sufficiently smooth in time.

The 4 elements of $V_{pq}$ represent the 4 possible correlations between the two pairs of output signals from the two stations of an interferometer.  These signals are usually labelled $X$ and $Y$ for
linearly polarized receptors, or $L$ and $R$ for circularly polarized receptors\footnote{A ``linearly polarized'' receptor is sensitive to linearly polarized radiation, e.g. a dipole.}:

\[ V_{pq} =
\left( \begin{array}{cc}
v_{XX} & v_{XY} \\
v_{YX} & v_{YY} 
\end{array} \right) 
\; \mbox{or} \;
\left( \begin{array}{cc}
v_{LL} & v_{LR} \\
v_{RL} & v_{RR} 
\end{array} \right)
\]

in which element $v_{YX}$ predicts the value of the correlation
between the $Y$ signal of station $p$ with the $X$ signal of station
$q$ etc.  

\subsection{\label{sec:implement-me}Implementing the RIME in MeqTrees}

\begin{figure}
\begin{centering}\includegraphics[width=\columnwidth]{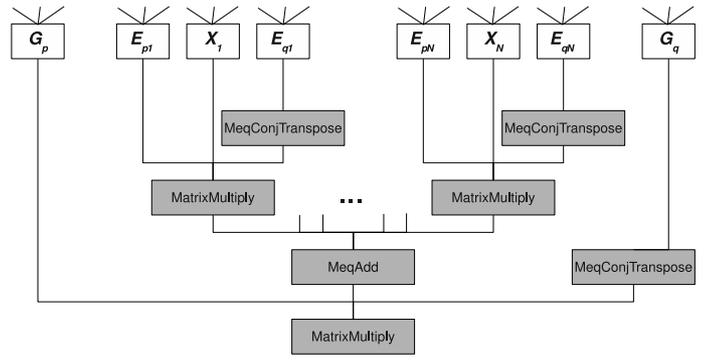}
\par\end{centering}
\caption{\label{fig:me}A subtree implementing the RIME given by eq.~(\ref{eq:rime}).}
\end{figure}

Implementing an equation like (\ref{eq:rime}) in MeqTrees is very straightforward. On the top level,  eq.~(\ref{eq:rime}) is just a small subtree composed of MeqAdd, MeqMatrixMultiply and MeqConjTranspose 
nodes (see Fig.~\ref{fig:me}). Such a subtree is constructed for every $p,q$ pair. Note that the children of the subtree -- nodes returning the constituent matrices of eq.~(\ref{eq:rime}), shown in lighter colour in the figure -- can themselves be represented by arbitrary subtrees. 

Figure~\ref{fig:me} captures the essense of the modularity and flexibility of MeqTrees. The ability to plug in  arbitrary subtrees (including arbitrary solvable parameters) to represent the source contributions $\vec X_k$ and the Jones matrices $\vec E_{pk}$ and $\vec G_p$ effectively allows for {\bf arbitrary parameterizations} of the sky and instrumental models. 

\subsection{\label{sec:LSM}The Local Sky Model}

The RIME predicts the visibilities observed by a particular radio telescope, given a particular source distribution. In eq.~(\ref{eq:rime}),  $\vec X_{k}$ is the intrinsic coherency that represents source $k$. The exact form of this matrix, and the corresponding subtree of Fig.~\ref{fig:me}, depends on the source model. 
In principle, $\vec X$ is a function of $u,v$ coordinates $\vec X(u,v)$ (which, per each baseline $pq$, are themselves a function of time and frequency) that is a Fourier Transform (F.T.) of the brightness distribution $\vec{\cal B}(l,m)$, relative to source center. In the case of a point source (i.e. a delta function), this is trivial:

\begin{equation}
\vec X =
\left( \begin{array}{cc}
I+Q & U+iV \\
U-iV & I-Q 
\end{array} \right) 
\; {\mathrm or}
\;
\left( \begin{array}{cc}
I+V & Q+iU\\
Q-iU & I-V 
\end{array} \right),
\label{eq:coh}
\end{equation}

depending on whether a linear (i.e. orthonormal) or circular polarization basis is used.\footnote{Some formulations include a factor of 1/2 in the definition of $\vec X$. See \citet{Smirnov:3C147} for a discussion of these issues. Note also that it is common to use $\vec B$ instead of $\vec X$, calling it the source {\em brightness} (and indeed, the brightness of a source at the phase centre is equivalent to its coherency.) In this paper we'll use $\vec X$, reserving $\vec B$ for the bandpass Jones, below.} 
For extended sources, more complicated forms of $\vec X(u,v)$ may be provided via their own subtrees. The following  forms have been implemented in MeqTrees at time of writing:

\begin{description}

\item[Gaussian components.] Slightly extended sources may be approximated by a two-dimensional Gaussian distribution in the $lm$-plane, as is done in the NEWSTAR package. The F.T. of this is a Gaussian in the $uv$-plane, which is provided by a simple subtree with flux, extent and orientation parameters.

\item[Images.] Most 2GC packages use images (i.e. a gridded $\vec{\cal B}(l,m)$ representation) as their standard sky model. For images, an FFT followed by degridding provides a computationaly effective way of estimating $\vec X(u,v)$. MeqTrees implements this approach via a combination of MeqFFTBrick and MeqUVInterpol nodes \citep{Abdalla:uvbrick}.

\item[Shapelets] are another way to efficiently model extended source structure. $\vec{\cal B}(l,m)$ is decomposed into shapelets in the $lm$-plane; this can be efficiently evaluated in the $uv$-plane.  The use of this in MeqTrees has been pioneered by \citet{Yatawatta:shapelets}.

\end{description}

The above source representations may be freely mixed, simply by plugging in different kinds of subtrees for the different $\vec X_k$ terms in Fig.~\ref{fig:me}. Note how this differs from the traditional 2GC view of a single ``sky image''. An image has the advantage of modelling arbitrary source structure, but it is limited by gridding errors (and distortions introduced by DDEs.) To maximize dynamic range, a mixed sky model may be required. Such a mixed model may consist of, e.g. point sources and shapelets for the brightest sources, and one or more images for the faint background. This is in principle straightforward to implement via different subtrees.

\begin{figure}
\begin{centering}\includegraphics[width=0.7\columnwidth,angle=-90]{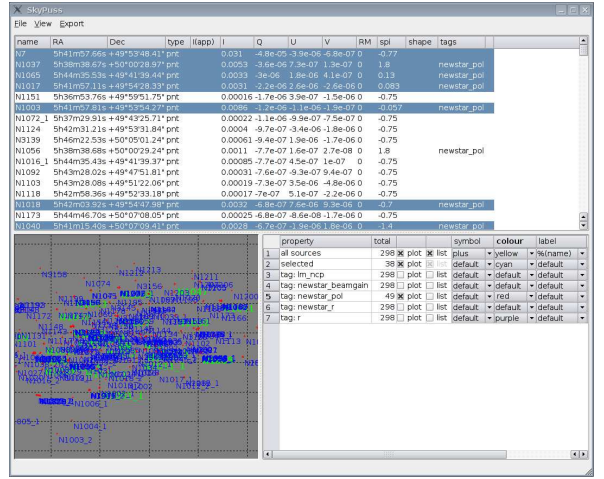}
\par\end{centering}
\caption{\label{fig:lsm}The user interface to the MeqTrees Local Sky Model}
\end{figure}

The definitions of the various sources that are relevant for a particular observation are stored in a Local Sky Model (LSM). MeqTrees provides an end-user tool for managing this information (see Fig.~\ref{fig:lsm}). Once the user has supplied an LSM, the relevant subtrees are constructed programmatically.

Last but not least, we should note that the parameters of the different source parameterizations are automatically functions of, e.g., frequency and time, and may in principle be solved for just like any other parameter in a tree (always provided the observational data is sufficient to constrain the problem, of course.) MeqTrees makes no distinction between instrumental and source parameters: it is possible to solve for {\em any} subset of RIME parameters.

\subsection{\label{sec:Jones}Jones matrices}

In eq.~(\ref{eq:rime}), the intrinsic LSM source coherency matrices $\vec X_{k}$ are ``corrupted''  
by means of 2$\times$2 Jones matrices. These are the real heart of the RIME. They represent the DDEs ($\vec E_{pk}$) associated with station $p$ and direction $l_k,m_k$, and DIEs ($\vec G_p$) associated with station $p$. 

$\vec E_{pk}$ and $\vec G_p$ are in principle themselves matrix products of a series of Jones matrices corresponding to individual physical effects.\footnote{Note that, because the signal path to each station is completely described by its own series of Jones matrices, the RIME is valid for arrays with very dissimilar stations, as is the case for LOFAR, and will probably be the case for the SKA.} Matrix multiplication does not commute, so the individual Jones terms must be placed into the equation in the correct order, corresponding to their physical order in the signal propagation path (but see below.) 

It is very useful to have a common letter-based nomenclature for the standard Jones matrices. Below we will give a by no means exhaustive list, with nomenclature mostly following \citet{JEN:note185}. Where appropriate, we will mention what form the Jones matrix usually takes. Three important forms are the {\em diagonal} matrix, the {\em rotation} matrix, and the {\em scalar} matrix:

\[\left( \begin{array}{cc}
d_{11} & 0 \\ 0 & d_{22}
\end{array} \right) 
\mbox{~~~~~~}
\mathrm{Rot}\,\phi = \left( \begin{array}{cc}
\cos\phi & -\sin\phi \\ \sin\phi & \cos\phi
\end{array} \right) 
\mbox{~~~~~~}
d \equiv \left( \begin{array}{cc}
d & 0 \\ 0 & d
\end{array} \right) 
\]

Note that diagonal or rotational form is subject to choice of basis. For example, a rotation in an orthonormal basis becomes a special kind of diagonal matrix in the circular polarization basis. (Below, we will assume an orthonormal basis unless otherwise noted.) Scalar matrices, on the other hand, are scalar regardless of basis. Within formulae, we shall use normal-weight capitals (e.g. $A$ as opposed to $\vec A$) to distinguish scalar matrices.

These three forms are important due to matrix commutation. Scalar matrices commute with everything, while diagonal matrices commute among themselves, as do rotations. This allows us to commute certain Jones terms around the RIME in order to derive new forms of the equation and/or for purposes of computational efficiency.

Common DDEs, in order of appearance in the signal path, are:

\begin{description}

\item[\bf K-Jones:] The phase term, accounting for the geometric delay and fringe stopping associated with station $p$ and direction $l,m$. This is a scalar matrix of the form $K_p = \exp{i(u_pl+v_pm+w_pn)}$, so it may be commuted to any part of the RIME (in fact, phase itself is a sum of contributions from different parts of the signal path, which are commuted together into the overall K-Jones.) Commuting $K_p$ and $K^\dagger_q$ next to each other, then multiplying them gives the Fourier Transform kernel -- it could be said that K-Jones is at the heart of all interferometery! 

\item[\bf Z-Jones:] Ionospheric phase and amplitude effects. The latter are usually small enough to be
ignored. The phase delay $\zeta$ has a known frequency dependence $(\propto\nu^{-1})$. $Z$ is a scalar matrix, and so may be commuted anywhere.\footnote{This is rather fortunate for LOFAR, because it means that the large ionospheric phase effects may be calibrated at a convenient point early in the process.} For narrow fields, $Z$ can be treated as a DIE (and, for calibration purposes, is absorbed in G-Jones, see below).

\item[\bf F-Jones:] Ionospheric Faraday rotation. This is a rotation matrix; the angle has a frequency $(\propto\nu^{-2})$ dependence. 

\item[\bf T-Jones:] Tropospheric phase delay and extinction, another scalar matrix. For narrow fields, $T$ can be treated as a DIE, and also absorbed in G-Jones during calibration. Alternatively, its values may be provided externally by water-vapour radiometers, as with mm-wave telescopes like ALMA. 

\item[\bf E-Jones:] The station beamshape (i.e. primary beam gain/phase in direction $l,m$). This is the most telescope-specific Jones matrix of them all, and the least well understood. For 2GC, it is usually assumed that E-Jones is time-independent and identical across stations, which means that it can be incorporated into the local sky model (in the form of apparent rather than intrinsic fluxes.) That this is not the case can severly limit imaging fidelity at instruments such as the VLA \citep{SB:imageplane}. Indeed, it can be argued that every radio telescope has or will have an E-Jones problem. 

\item[\bf P-Jones:] Projection matrix, corresponding to the projected position angle of the two receptors of a station on the sky. For dishes with narrow FoVs, this is a DIE, and is a simple rotation (constant for equatorial mounts, and offset by the time-variable parallactic angle for alt-az mounts.) For a horizontal dipole array like a LOFAR station, $\vec P$ becomes a more complicated expression (which can also be incorporated into an E-Jones model.)

\end{description}

Commonly used DIEs are:

\begin{description}

\item[\bf D-Jones:] ``On-axis'' polarization leakage\footnote{In 2GC practice, D-Jones is the  leakage in the direction of the dominating source, which is usually at the field centre.} between the two receptors. This may be parameterized in terms of dipole orientation error (for linear receptors) and ellipticity. Note that the concept of ``leakage'' itself is a carry-over from 2GC packages, and is only a first-order approximation to the rather complicated polarization behaviour that can be more generally described by using proper E-Jones models. D-Jones is usually an almost-unity matrix, with small non-zero off-diagonal terms. 

\item[\bf G-Jones:] Complex gain, the staple of 2GC. Nominally, this corresponds to the electronic gain of the receivers, but for calibration purposes it cannot always be distinguished from the Z- and T-Jones terms, and so ends up subsuming all three effects. It is a diagonal matrix (unless electronic cross-talk is also incorporated, in which case the off-diagonal terms take on small non-zero values) with rapid variation in time, but little to none in frequency.

\item[\bf B-Jones:] Electronic bandpass. This is a diagonal matrix like G-Jones, but it has considerable structure in frequency, and only slow (if any) variation in time. Since D-Jones varies on similar timescales, it may be useful to combine the two into a full $2\times2$ matrix. 
\end{description}

\subsection{Simulation vs. calibration}

Real-life applications of the RIME in MeqTrees have, to date, fallen into two broad categories, {\em simulation} and {\em calibration}. 

Simulation has become an increasingly important field in the past decade, due to the large number of new radio telescopes being designed and built. For simulation, the RIME (plus an optional noise term) predicts the output of a [real or theoretical] telescope observing a model sky. Given a sky model, and a set of Jones matrices for the DDE and DIE components (see e.g. Fig.~\ref{fig:cattery}), MeqTrees constructs a set of per-baseline trees corresponding to a RIME such as that in eq.~(\ref{eq:rime}), and evaluates them for a series of times and frequencies specified by a Measurement Set (MS). The resulting simulated visibility data is then written out to the MS. 

Calibration is, essentialy, model fitting, as described in Sect.~\ref{sec:solving}. The RIME is used to predict model visibilities (similar to the simulation case), but the outputs of the subtrees are then treated as the model functions ${M^{(pq)}}$ of Sect.~\ref{sec:solving}, and fitted (by solving for their parameters) to the observed data $D^{(pq)}$, which is read from an MS. The resulting residuals are also written out to the MS. If the parameters being solved for are the complex station gains (the G-Jones term), this procedure is the equivalent of traditional 2GC selfcal. However, since MeqTrees in principle allows for arbitrary parameterizations, the same approach can be used to solve for, e.g., coefficients of beam models, ionospheric models, etc. 

These two applications of the RIME are superficially similar, in that in both cases the RIME is used to predict model visibilities. However, the kinds of Jones matrices employed can be significantly different. For simulations, we are usually interested in a realistic representation of the underlying physics, so we use some of the Jones terms listed above, as relevant to a particular simulation. For example, a simulation may usefully incorporate separate Z-Jones, T-Jones and G-Jones terms, employing different numerical models for their matrix elements. During calibration, on the other hand, we may be unable to solve for these effects separately, since they all add up into one rapidly varying per-station phase term (assuming a narrow frequency band and FoV, where the different frequency behaviour of Z-, T- and G-Jones cannot be distinguished, and Z- and T-Jones become direction-independent.) Nor do we need to, since it is only the overall phase term that we need to calibrate in order to image the data properly. Therefore, for calibration purposes, all three effects can be captured by a single G-Jones term with solvable (complex) diagonal elements. 

We call such forms of the RIME {\em phenomenological}, since they are meant to provide sufficient degrees of freedom (i.e. solvable parameters) to capture the {\em effect} of the instrument on observed data, with little regard to the underlying physics. Here's an example of a real-life phenomenological RIME for polarization calibration of the WSRT:

\begin{equation}
\vec V_{pq} = \vec B_{p} \vec G_{p} \vec P \left( \sum^{N}_{k=1} \vec E_k K_{pk} \vec X_k K^\dagger_{qk} \vec E^{\dagger}_k \right) \vec P^{\dagger} \vec G^{\dagger}_{q} \vec B^{\dagger}_{q}
\label{eq:rime:bg}
\end{equation}

Here, $\vec B_p$ is a solvable full $2\times2$ matrix to capture bandpass and leakage (highly variable in frequency, but only slowly variable in time), $\vec G_p$ is a solvable diagonal matrix to capture rapidly variable amplitude and phase effects (no variation in frequency, rapid variation in time), $\vec P$ is a dipole orientation matrix (known, constant, and same for all stations), $\vec E_k=\vec E(l_k,m_k)$ is an apriori primary beam model (same for all stations), and $K_p$ is the usual phase term. 

At very high ($>10^5$) dynamic ranges, short-term low-level instability of the WSRT bandpass makes it impossible to separate the $\vec B$ and $\vec G$ solutions, so a simpler form of the RIME may be used instead. This form is roughly equivalent to per-channel selfcal in 2GC terminology:

\begin{equation}
\vec V_{pq} = \vec G_{p} \vec P \left( \sum^{N}_{k=1} \vec E_k K_{pk} \vec X_k K^\dagger_{qk} \vec E^{\dagger}_k \right) \vec P^{\dagger} \vec G^{\dagger}_{q} 
\label{eq:rime:g}
\end{equation}

Here, $\vec G_p$ is solvable, with rapid variation in frequency and time on the diagonal, and slow 
variation in time on the off-diagonal. Another version of this equation was used by \citet{Smirnov:3C147} to produce the result in Fig.~\ref{fig:3c147} (see Sect.~\ref{sec:Results}):

\begin{equation}
\vec V_{pq} = \vec G_{p} \vec P \left( \sum^{N}_{k=1} \vec{\Delta E}_{pk} \vec E_k K_{pk} \vec X_k K^\dagger_{qk} \vec E^{\dagger}_k \vec{\Delta E}^\dagger_{qk} \right) \vec P^{\dagger} \vec G^{\dagger}_{q} 
\label{eq:rime:de}
\end{equation}

Here, {\em differential gain} $\vec{\Delta E}_{pk}$ is a phenomenological Jones term capturing the source-dependent complex gain variations (which can be due to any combination of physical effects). It is diagonal, and solvable on large time and frequency scales.

\subsection{RIME and software modularity}

Early thinking about implementations of the RIME in software \citep{JEN:note185} tended towards a ``one equation to rule them all'' doctrine. A particular sequence of Jones terms was chosen and carefully elaborated on, with the implicit expectation that it would be appropriate to all telescopes and scenarios, if only all the required Jones matrices were properly implemented. This thinking also had a strong influence on the AIPS++/CASA calibration modules.

Our subsequent work on the subject has convinced us that a more flexible approach is vital. The discussion in the previous section suggests that no specific set of Jones terms can be appropriate to all cases -- not even if we restrict ourselves to one relatively simple instrument such as the WSRT, as is clear from eqs. (\ref{eq:rime:bg}--\ref{eq:rime:de}). The need to build {\em arbitrary} measurement equations drove the design of MeqTrees from the beginning.

Fortunately, the structure of the RIME itself encourages just such a flexible and modular approach. Any specific knowledge about an instrument can be encoded in the form of Jones matrices, and all Jones matrices work the same way. This has the very considerable advantage that each instrumental effect can be modelled in isolation, while it will still be taken into account correctly within the RIME. {\em This is a far cry from the intractable and approximate mathematical expressions in older data reduction packages, which are often not implemented explicitly, but scattered in poorly documented bits and pieces throughout the code.}

\begin{figure}
\begin{centering}
\includegraphics[width=0.9\columnwidth,angle=0]{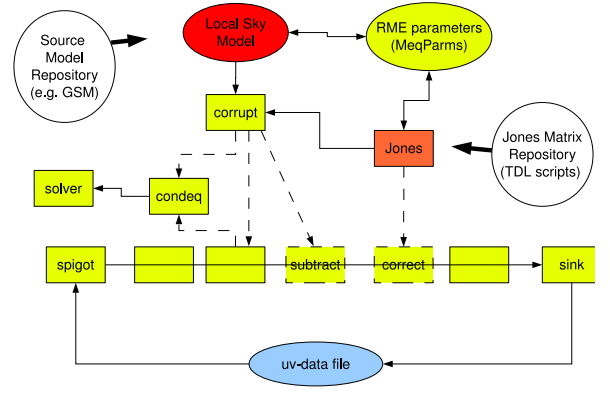}
\par\end{centering}
\caption{\label{fig:procscript}Schematic structure of a typical TDL 
processing script. Thanks to the RIME, with its Jones matrices and Local 
Source Model, such scripts are highly modular, and may quickly be adapted 
to different telescopes and experiments. This greatly facilitates 
collaboration and rapid experimentation. See also the text.}
\end{figure}

\begin{figure}
\begin{centering}
\includegraphics[width=0.9\columnwidth,angle=0]{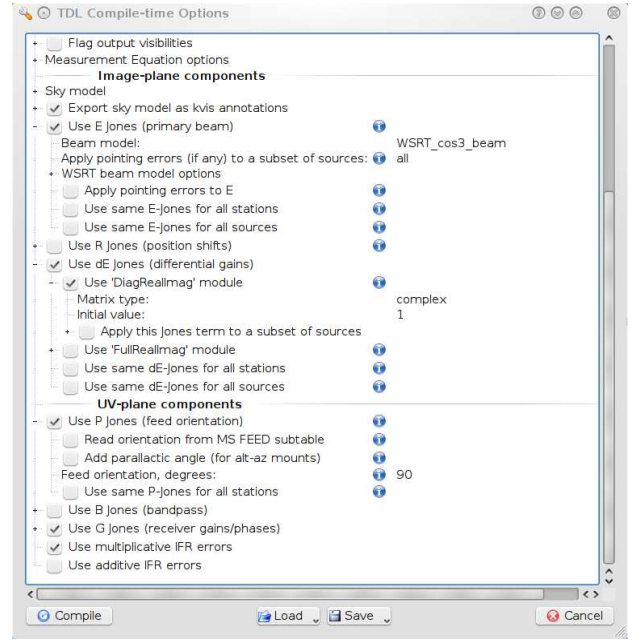}
\par\end{centering}
\caption{\label{fig:cattery}The Options GUI of a typical processing script. A number of Jones
modules may be selected for inclusion in the tree. Each Jones module can implement its own 
custom option set, which is displayed in submenus.}
\end{figure}

Figure~\ref{fig:procscript} shows schematically how the modularity of the RIME can be exploited to generate a set of generic processing scripts that can be adapted for use with any radio telescope. Rather than being written in basic TDL from scratch, these scripts are based on a set of generic Python frameworks called the Cattery, which ships with MeqTrees 1.1. The user selects a Cattery script for a particular operation (e.g. simulation, or visualization, or some sort of calibration), and perhaps customizes it by changing a few lines of Python code. The user then selects a suitable sequence of Jones modules from a ``Jones repository''.\footnote{At time of writing, a number of standard Jones modules were included in the MeqTrees 1.1.1 binary release. More modules can be checked out manually from a designated area on our Subversion server. We intend to set up a more formally structured repository as more diverse Jones modules are implemented.} Finally, the user selects a Local Sky Model, for which a number of model formats are supported. All these modules are free to define their own custom processing options (solvable parameters, etc.) MeqTrees automatically extracts the relevant option set from the selected modules, and presents it to the user via a GUI (Fig.~\ref{fig:cattery}). When used in batch mode, it can load these options from configuration files.

\section{Some results\label{sec:Results}}

\begin{figure}
\begin{centering}\includegraphics[width=\columnwidth]{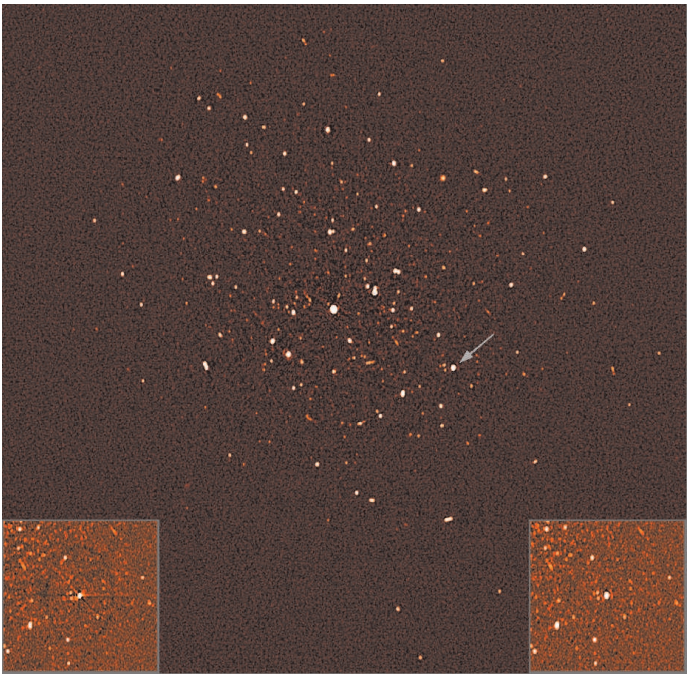}\par\end{centering}
\caption{\label{fig:3c147}This WSRT 21cm image of the
field around the bright radio source 3C147 is virtually noise-limited,
and has a dynamic range of 1.6 million. This dynamic range, achieved
by \citet{deBruyn:3C147} using regular selfcal with the NEWSTAR
package, is high enough to clearly show ring-like artifacts around
moderately bright off-axis sources (see left inset) which are caused
by DDEs. We used MeqTrees to apply differential gain solutions to the
original data (with a sky model built up by de Bruyn et al. during
their NEWSTAR reduction), which completely eliminated the artifacts
(right inset).}
\end{figure}

\begin{figure}
\begin{centering}\includegraphics[width=\columnwidth]{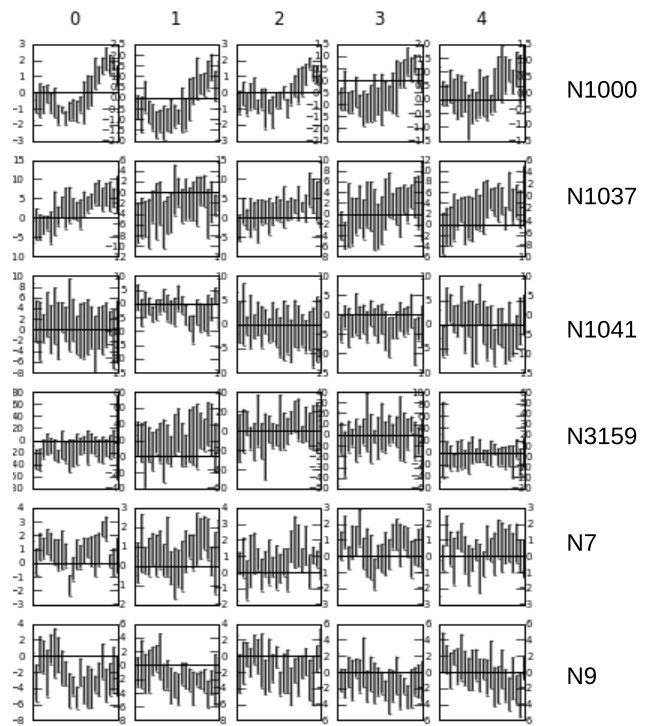}\par\end{centering}
\caption{\label{fig:diffgains}
Differential phase solutions in the direction of 6 different sources (per 5 antennas), 
as a function of time. The integration time is 30 minutes. Note that the S/N is large 
enough to show slow variations.}
\end{figure}

\citet{Smirnov:beat-NEWSTAR2,Smirnov:3C147} has used MeqTrees to address DDEs in high-dynamic range WSRT observations of the 3C147 field. WSRT has been the ``world champion'' in dynamic range for the last few decades due to its extremely favourable design characteristics, and in particular comparatively benign DDEs, but at extreme dynamic ranges (and even not so extreme, in some observational modes) DDEs do become a problem. The result in Fig.~\ref{fig:3c147} shows how a modified form of the RIME may be used to address it. This image has a virtually noise-limited dynamic range of 1.6 million. It is an improvement over the image obtained with NEWSTAR by \citet{deBruyn:million,deBruyn:3C147} with the same data, primarily because it corrects for DDEs in the map by solving for differential gains (eq.~\ref{eq:rime:de}). With WSRT, time-variable DDEs manifest themselves as faint ring-like structures around off-axis sources (see left inset of Fig.~\ref{fig:3c147}), which cannot be deconvolved. Differential gains eliminate these structures (see right inset). 
Figure \ref{fig:diffgains} shows the differential phase solutions for 5 of the 14 WSRT telescopes as a function of time in the direction of six moderately bright sources in the field. The values are relative to the phase solution in the direction of the dominating source (3C147, 22 Jy). With an integration time of 30 minutes, the S/N of the differential phases is clearly large enough to detect slowly varying real effects. Interpreting these is another matter, and will be discussed further in a separate paper \citep{Smirnov:3C147}. 

\citet{Yatawatta:lofar} has been using MeqTrees to make the first all-sky images with initial
LOFAR stations. LOFAR observations are subject to complex DDEs caused by the ionosphere and time-variable primary beam patterns, so its calibration is a challenge, with many alternative approaches being (and waiting to be) explored. MeqTrees, with its rapid experimentation capability, has proven to be a perfect vehicle for this.

On the simulations side, the ability to implement arbitrary RIMEs has allowed MeqTrees to be used for simulation of new and unusual telescope designs. \citet{Willis:sims2,Willis:sims1} has been using MeqTrees to simulate interferometers composed of dishes with Focal Plane Arrays (FPA), with a particular emphasis on studying the effects caused by rotating beam patterns (as in an alt-az mount without a derotator) and polarimetric fidelity. Mevius and Van Bemmel have been using MeqTrees to develop an ionospheric simulations framework called LIONS \citep{Anderson:ionosphere}.

\section{\label{sec:Conclusions}Conclusions}

MeqTrees raises the art of instrumental modelling to the level where
user-developers can concentrate on the physics of the problem, while
the complex numerical machinery, e.g. for solving for arbitrary
subsets of parameters, is hidden ``under the hood''. In the special case
of radio astronomy, the correct treatment of various instrumental
effects is greatly facilitated by the elegant matrix formalism of the
Measurement Equation of a generic radio telescope (RIME). The latter is
well on its way to becoming the new Common Language of radio astronomy.

The present collection of node classes offers basic functionality,
with bias towards radio astronomy (see Appendix
\ref{sec:Available-MeqNode-classes}). There are various TDL (Python)
frameworks to help the user in building complex trees; these in fact
evolve much more rapidly than the binary release cycle. We also intend
to offer a MeqWizard tool to help both novices and experts to find
their way in the multiverse of possibilities.  The MeqTrees kernel is
robust and efficient, and has been tested thoroughly on real data (see
Sect. \ref{sec:Results}.)

MeqTrees is (slowly) beginning to find its place, propelled by the
increasingly urgent need for 3GC simulation and calibration software
for radio astronomy. It plays its designated role as pathfinder for
LOFAR calibration, as illustrated by impressive all-sky LOFAR
images \citep{Yatawatta:lofar}. It has been used as the main
education tool in several international workshops, to train the the
new generation of radio astronomers in the use of the RIME and 3GC. 

MeqTrees is freely distributed under the terms of the GNU General
Public License. A stable binary release (version 1.1.1 at time of writing) is available.
This is shipped as binary packages for the major Linux distros, so installation is relatively painless (while users of unsupported
platforms always have the option of building from source.) A Mac OSX version has been tested, but is not (yet) part of the binary release. MeqTrees is natively multi-threaded to take full advantage of multi-core machines common today. An experimental MPI-based cluster version was developed jointly with Oxford Astrophysics and OeRC, and is currently being tested by Tony Willis at DRAO (Canada.)

For further information on downloading and installing the software, please  refer to the MeqTrees Wiki: {\tt http://www.astron.nl/meqwiki}. You can also join the MeqTrees forum hosted at UCL: {\tt https://great08.projects.phys.ucl.ac.uk/meqtrees/}.

\begin{acknowledgements}

Up to now, the following people have made direct contributions to
MeqTrees, either by implementing and testing nodes, or by writing and
exercising TDL scripts: Sarod Yatawatta, Tony Willis, Maaijke Mevius,
Ger van Diepen, Ronald Nijboer, Filipe Abdalla, Rob Assendorp, Ilse
van Bemmel, Ian Heywood, Hans-Rainer Kloeckner, Rense Boomsma, Michiel
Brentjens, Joris van Zwieten, Alessio Magro. Many more should follow,
and will be acknowledged, as our collaborative network grows.

Others have contributed to the distribution of MeqTrees and its
operation on different platforms: Chris Williams, Stef Salvini, Mike
Sipior, James Anderson and George Heald. Ger de Bruyn has provided
data, models, advice, and a wealth of inspirational ideas.
Helpful comments were extracted from Wim Brouw. MeqTrees uses various modules from
AIPS++/CASA, particularly those written by Ger van Diepen, Wim Brouw
and Tim Cornwell. Ger van Diepen and Malte Maquarding have also been
instrumental in making the CASA installation more robust. For the
application of DDEs we have been influenced by the work of Sanjay
Bhatnagar and Steve Gull. The MeqTrees binary package repository is kindly 
hosted by the Oxford e-Research Centre, and operated by Chris Williams.
The MeqTrees Forum website is kindly hosted by University College London, 
and operated by Julien Girard of the Observatoire de Paris.

The authors are particularly grateful for the faith and patience of a
succession of ASTRON managers over the years: Marco de Vos, Mike
Garrett, Ronald Nijboer, Arnold van Ardenne, Kjeld van der Schaaf,
Harvey Butcher.  And finally, the very substantial support of Steve
Rawlings (Oxford Astrophysics), Peter Dewdney and Sean
Dougherty (NRC DRAO) and Anne Trefethen (Oxford e-Research Centre) 
has made all the difference.

\end{acknowledgements}

\bibliographystyle{aa}
\bibliography{15013}
 
\appendix

\section{\label{sec:Available-MeqNode-classes}Available node classes}

Perhaps the most concise description of the capabilities of MeqTrees
for the discerning user is an overview of the node classes that are
available already, and the ones that are desirable in the near future:

\begin{description}
\item[Leaf nodes:] Constant, Parm, Freq, Time, Grid, GaussNoise, RandomNoise, Spigot, FITSImage (and several
other FITS interface nodes)

\item[Unary operations:] Exp, Log, Abs, Invert, Negate, Sqrt,
Pow2(8), Sin, Cos, Tan, Acos, Asin, Atan, Cosh, Sinh, Tanh, Norm,
Arg, Real, Imag, Conj, Ceil, Floor, Identity. See also Sect. \ref{sec:tensors}.

\item[Binary operations:] (two children): Subtract, Divide, Pow, Mod,
ToComplex(real,imag), Polar(ampl,phase). See also Sect. \ref{sec:tensors}.

\item[Accumulation:] (one or more children): Add, Multiply, WSum,
WMean. The last two need a vector of weights. See also Sect. \ref{sec:tensors}.

\item[Reduction nodes:] Sum, Mean, Product, StdDev,
Rms, Min, Max, NElements. These reduce a result along selected
axes. The default is all axes, in which case the result is a scalar.
See also Sect. \ref{sec:tensors}.

\item[Tensor operations:] Composer, Selector, Paster. Tensor nodes are nodes
with multiple vellsets in their Result (see Sect. \ref{sec:tensors}). 

\item[Matrix operations:] Transpose, ConjTranspose, MatrixMultiply,
MatrixInvert22. The latter operates on 2x2 matrices only, which is
sufficient for the RIME (see Sect. \ref{sec:RIME}).

\item[Flow control:] ReqSeq, ReqMux, Sink, VisDataMux. They regulate
the order in which their children get Requests, and which Result to
pass on. They also synchronize the flow of Requests and Results in
parallel trees.

\item[Domain Control:] ModRes, Resampler, CoordTransform. ModRes modifies
the Request before it is passed on, Resampler modifies the Result
itself. CoordTransform modifies the Request passed to its second child.

\item[Flagging:] ZeroFlagger, MergeFlags (see Sect.
\ref{sec:flags}). 

\item[Solving:] Condeq, Solver, Parm (see Sect. \ref{sec:solving}).
For the moment, MeqTrees offers only a Levenberg-Marquard non-linear
solver. 

\item[Visualization]: All nodes, DataCollect, Inspector(=Composer)
(see Sect. \ref{sec:Visualization}).

\item[Coordinates:] (mostly radio astronomy): UVW, LMN(radec,radeq0),
AzEl(radec,xyz), RaDec(azel,xyz), LMRaDec(lm), ObjectRaDec(name),
LST(domain,xyz), ParAngle(radec,xyz), LongLat(xyz). The vector {\em xyz}
is an Earth position in IRTF coordinates.

\item[Transforms:] FFTBrick, UVInterpol (collectively known as UVBrick)
(see Sect. \ref{sec:RIME}).

\item[User-definable nodes:] PyNode, Functional, PrivateFunction.
These allow the user to insert his own function, written in Python
or C++, to read the Results from one or more child nodes, and to generate
a customized Result. 

\end{description}

A full description of these nodes is outside the scope of this paper.
The present collection has an obvious bias towards radio astronomy.
This may change as MeqTrees is used in other application areas. We
envisage a core collection of nodes that offer basic functionality,
surrounded by collections of more specialised nodes (and Python frameworks)
for use in specific areas. The latter should be mostly contributed by users
-- we strive towards an Open Source developement model where everybody contributes,
while a small core team keeps the mainline development on track. 
Some of the contributed nodes will eventually find their way into the core collection,
while some of the present nodes will be moved out. In the meantime,
the various types of user-definable nodes (such as the PyNode) allow experimentation
with new node classes before they are implemented as a regular C++
node class. Obviously, we will have to solve the technical problem
of linking such contributed nodes into MeqTrees with a minimum of
fuss.

\end{document}